\documentclass[12pt]{article}

\usepackage{amsmath}
\usepackage{graphicx}
\usepackage{amssymb}
\usepackage{amsfonts}
\usepackage{latexsym}

\def\beq{\begin{eqnarray}}
\def\eeq{\end{eqnarray}}

\renewcommand{\vec}[1]{{\bf #1}}

\def\ka{\kappa}

\begin{document}
\begin{titlepage}
\title{Bouncing Universes in Scalar-Tensor Gravity Around Conformal Invariance}
\author{B. Boisseau$^1$\thanks{email:bruno.boisseau@lmpt.univ-tours.fr}, 
H. Giacomini$^1$\thanks{email:hector.giacomini@lmpt.univ-tours.fr},
D. Polarski$^2$\thanks{email:david.polarski@umontpellier.fr},
\hfill\\
$^1$Universit\'e de Tours, Laboratoire de Math\'ematiques et Physique Th\'eorique,\\
CNRS/UMR 7350, 37200 Tours, France\\
$^2$Universit\'e Montpellier \& CNRS, Laboratoire Charles Coulomb,\\
UMR 5221, F-34095 Montpellier, France\\ }

\pagestyle{plain}
\date{\today}

\maketitle

\begin{abstract}
We consider the possibility to produce a bouncing universe in the framework of 
scalar-tensor gravity when the scalar field has a nonconformal coupling to the Ricci 
scalar. We prove that bouncing universes regular in the future with essentially the 
same dynamics as for the conformal coupling case do exist when the coupling deviates 
slightly from it. This is found numerically for more substantial deviations as well. 
In some cases however new features are found like the ability of the system to leave 
the effective phantom regime. 
\end{abstract}
\end{titlepage}


\section{Introduction}
Bouncing universes have attracted a lot of interest for various reasons. 
While on one hand they crucially modify the evolution of the universe by 
avoiding the initial singularity, they are also interesting to study from a mathematical 
point of view.
A bouncing universe with a nonzero measure set of initial conditions can be obtained 
with a massive scalar field in a closed FLRW universe \cite{Star78} where the curvature 
singularity is generically moved to the past. 
Non-singular solutions are degenerate (i.e. they exist only for a set of initial conditions 
which is of measure zero) \cite{Page84}, see also \cite{Ka98}, while a bounce with positive 
spatial curvature requires severe fine tuning of initial conditions before the contraction 
stage \cite{Star78}, \cite{GT08}.
Spatially-flat FLRW non-degenerate bouncing universes have been built outside general 
relativity like theories with scalar~\cite{DSNA11,BS13} or tensor ghosts, loop quantum 
gravity (see e.g. \cite{APS06}) or gravity described by an effectively non-local 
Lagrangian~(see also \cite{QECZ11}, \cite{ESV11}, \cite{CEB12} and \cite{BP14}, \cite{BP16} 
for recent reviews).
In the framework of general relativity (GR) this requires the presence of some component of 
the phantom type around the bounce. In this sense interest in bouncing universes has an 
overlap with the field of dark energy (DE), where components of the phantom type are often 
considered. In particular, most modified gravity theories can produce an effective component 
of the phantom type, a property often put forward in favour of dark energy models outside 
general relativity (GR). Let us note that recent data do actually support the possibility of 
an effective DE component of the phantom type (see e.g. \cite{Pl13}). 
In this work we consider some general class of scalar-tensor theories of gravity, 
a well understood and widely studied contender of GR. It is known that 
these theories can violate the null-energy condition \cite{BEPS00},\cite{T02},\cite{GPRS06}. 
However, the construction of a concrete regular nondegenerate spatially flat bouncing model 
is a highly nontrivial challenge. 
It came therefore as a surprise that this could be achieved in a simple model with a 
conformally coupled scalar field and a negative potential \cite{BGPS15}. Moreover the problem 
turned out to be completely integrable and exact analytical solutions were found. Translating 
this solution in the Einstein frame \cite{BGP15} gives an integrable minimally coupled scalar 
field with an inverted double-well potential, also found without relation to bouncing solutions 
in the Jordan frame in \cite{BCST12}, while the viability bounds $F=0$ correspond to either 
a Big Bang or a Big Crunch \cite{BGP15}. These bouncing solutions were extended to non-flat 
geometries in \cite{KPTVV15}. 

While its importance was clear regarding the possibily to derive exact solutions, the 
relevance of the conformal coupling for the existence and the dynamics 
of bouncing solutions remained unclear. The importance of conformal symmetry has been 
recognized and used in many physical problems (see e.g. \cite{tH15}, \cite{Boy16}, 
\cite{Rub09}, \cite{MTZ03}, \cite{HPP06}). This motivates us even more to assess its 
relevance here. Also, the conformally invariant bouncing model found rested on a background 
satisfying $\dot{H}>0$ forever after the bounce, which can be seen as a problem if one is 
willing to build a realistic cosmological scenario with it. It is therefore interesting to 
see whether a richer behaviour is obtained with a slight generalization of our model.     
We will consider minimal breaking of conformal invariance by studying a wider class of 
couplings (see \cite{H15} for a similar study with other motivations) while leaving the 
potential unchanged, and investigate in how far the bouncing solutions, if they still exist, 
depart from the bouncing solutions found earlier with conformal coupling. 
  
\section{The conformally invariant bouncing model}
We consider a universe where gravity is described by a scalar-tensor model. The 
Lagrangian density in the Jordan frame of the gravitational sector is given by 
(see e.g. \cite{EP00})
\begin{equation*}
\label{L}
L = \frac{1}{2}\left( F(\Phi)R - Z(\Phi)~g^{\mu\nu}\partial{\mu}\Phi\partial{\nu}\Phi   
                                            - 2 U(\Phi) \right)~.
\end{equation*}
We can take $Z=1$ or $Z=-1$, corresponding physically to 
$\omega_{BD}>0$ or $\omega_{BD}<0$. 
For $\omega_{BD}<0$, the theory is ghost-free provided $-\frac{3}{2}<\omega_{BD}<0$.  
We consider further spatially flat FLRW universes with metric $ds^2=-dt^2+a^2(t)d\vec{x}^2$ 
yielding the following modified Friedmann equations 
\beq
3 F H^2  &=&  \frac12 Z~\dot{\Phi}^2 - 3 H \dot{F} + U ~, \label{Fr1}\\
-2 F \dot{H} &=& Z~\dot{\Phi}^2 + \ddot{F} - H \dot{F}~, \label{Fr2}
\eeq
with $H\equiv \frac{\dot{a}}{a}$. Here and below a dot, resp. a prime, stands for the 
derivative with respect to $t$, resp. to $\Phi$. 
Equations \eqref{Fr1},\eqref{Fr2} contain the equation of motion of $\Phi$
\beq
Z~(\ddot{\Phi} + 3 H \dot{\Phi}) - 3 F'(\dot{H}+2 H^2) + U' = 0~. \label{ddPhi}
\eeq
We study the specific model
\beq
Z F &=& -\xi \Phi^2 + \kappa^{-2}, \label{F}\\
Z U &=& \frac{\Lambda}{\kappa^2} - c \Phi^4 ~.\label{U}
\eeq
The equations \eqref{Fr1} -- \eqref{ddPhi} then become 
\beq
3\left( -\xi\Phi^2 + \ka^{-2} \right) H^2 &=& \frac12 \dot{\Phi}^2 + 6H \xi \Phi \dot{\Phi} +
                                   \Lambda \ka^{-2} - c \Phi^4~,\label{Fr1xi}\\
-2\left( -\xi\Phi^2 + \ka^{-2} \right) \dot{H} &=& \dot{\Phi}^2 - 2\xi \dot{\Phi}^2 - 
            2\xi \Phi \ddot{\Phi} + 2H \xi \Phi \dot{\Phi} ~,\label{Fr2xi}
\eeq
\beq
\ddot{\Phi} + 3H \dot{\Phi} + 6\xi \Phi (\dot{H}+2H^2) - 4c\Phi^3 = 0~.
\label{ddPhixi}
\eeq
We emphasize that if we put all terms of eq.\eqref{Fr1xi} on the same side we obtain a 
vanishing first integral of eq.\eqref{Fr2xi}-\eqref{ddPhixi}. If 
$\xi =\frac16$, we recover the system studied in \cite{BGPS15} for which exact bouncing 
universes regular in the future were found, and equation \eqref{ddPhi} corresponds to
a massless scalar field conformally coupled to Einstein gravity with an interaction 
potential $-c\Phi^4$. We take $Z=1$ because for $Z=-1$ only one initial condition 
$\Phi_0\equiv \Phi_{0,cr}$ yields a viable solution. The solutions found were regular in the 
future in the usual sense that they do not diverge but also because they satisfy $F>0$ at 
all times after passing the bounce. For 
convenience we locate the bounce at $t=0$ with $H(t=0)\equiv H_0=0$ (we will adopt a similar 
notation for quantities at $t=0$). The allowed interval of initial values $\Phi_0$ is given 
by $\Phi_{0,min}\le \Phi_0 < \Phi_{max}$. The potential $U$ vanishes at $\Phi_{0,min}$ while $F$ 
vanishes at $\Phi_{max}$. We note from \eqref{Fr1xi} that the field 
velocity at the bounce is given up to a sign by the value of $\Phi_0$. 
A family of solutions with non-zero measure in the set of initial conditions was found with 
initial values between some critical value 
$\Phi_{0,cr}$ and $\Phi_{max}$ (and $\dot{\Phi}_0<0$) and tending asymptotically to zero in 
a monotonous way.
Remarkably, $H(t)$ was found to couple to a first integral of $\Phi$ only. So it is the same 
function of time $H = \sqrt{ \frac{\Lambda}{3} } \tanh \Big[2 \sqrt{\frac{\Lambda}{3}}~t\Big]$ 
irrespective of the specific dynamics of $\Phi$. 
The unique solution starting precisely at $\Phi_{0,cr}$ tends asymptotically to some 
non-zero value $\tilde\Phi\equiv (\frac{\Lambda}{6c})^{\frac{1}{2}}$.     
All solutions starting at $t=0$ in the interval $\Phi_{0,min}\le \Phi_0 < \Phi_{0,cr}$ were found 
to diverge in a finite time. 
While conformal invariance is obviously lost for $\xi\ne \frac16$, 
we will next show that bouncing solutions regular in the future exist in that case too. 

\section{Bouncing solutions for $\xi\ne \frac16$}

Let us first introduce the parameter $\epsilon$ defined as follows
\beq
\xi=\frac16 (1 + \epsilon)~.\label{xi}
\eeq 
Clearly, $\epsilon=0$ corresponds to conformal invariance. The condition $F>0$ 
holds for $\Phi<\Phi_{max}$, viz.
\beq
\Phi < \Phi_{max} = \frac{\sqrt{6}}{\kappa} \frac{1}{\sqrt{1 + \epsilon}}~, 
~~~~~~~~~~~~~~~~~~~~~~~~~Z=1~. \label{Phimax}
\eeq
Moreover, from \eqref{Fr1} we must have at the bounce 
$\Phi_0\ge \Phi_{0,min}=\left( \frac{\Lambda}{\ka^2 c}\right)^{\frac14}$. 
Note that $\Phi_{0,min}$ does not depend on $\epsilon$. Therefore the existence 
of a nonvanishing interval of initial values leading to regular bouncing solutions requires 
\beq
\sqrt{1 + \epsilon} < \Phi^{-1}_{0,min}~\Phi^{(0)}_{max}\equiv \sqrt{1 + \epsilon_{max}}~, 
\label{intZ1}
\eeq
where we have defined $\Phi^{(0)}_{max}\equiv \Phi_{max}(\epsilon=0)$, a useful notation that 
will be extended to all quantities. We have further introduced in \eqref{intZ1} the 
quantity $\epsilon_{max}$ 
\beq
\epsilon_{max} = \frac{6}{\kappa}~\left( \frac{c}{\Lambda} \right)^{\frac12} - 1~.\label{epsmax}
\eeq
The initial conditions of our system are completely defined from $\Phi_0$. Indeed from 
\eqref{Fr1} or \eqref{Fr1xi}, $\dot{\Phi}_0$ is fixed (up to a sign) once $\Phi_0$ is given. 
Moreover for a given $\Phi_0$, the same value $\dot{\Phi}_0$ is obtained independently of 
$\epsilon$. We will use this property later. 
It is also obvious that the allowed interval of initial values for $\epsilon > 0$ is 
smaller than the interval for $\epsilon = 0$ while it is larger for $\epsilon < 0$. 

We write further equations \eqref{Fr2xi}, \eqref{ddPhixi} in a form suitable for a dynamical 
system analysis
\beq
\dot{H} &=& - \frac{\kappa^2\left( (2 - \epsilon)y^2 + 4(1 + \epsilon)H y\Phi + 
 2(1 + \epsilon)^2\Phi^2H^2 - 4 c(1 + \epsilon)\Phi^4 \right)}
{6 + \epsilon (1 + \epsilon)\kappa^2 \Phi^2} ~,              \label{dyn1}\\
\dot{\Phi} &=& y~,                                           \label{dyn2}\\
\dot{y} &=& \frac{ H y \left( -18 + ( 4 + 5\epsilon + \epsilon^2)\kappa^2\Phi^2 \right) + 
           H^2\Phi \left( -12(1 + \epsilon) + 2(1 + \epsilon)^2 \kappa^2 \Phi^2 \right)} 
{6 + \epsilon (1 + \epsilon)\kappa^2 \Phi^2} \nonumber \\
     &+&  \frac{\Phi \left( (2 + \epsilon - \epsilon^2 )\kappa^2 y^2 + 
         4c ( 6\Phi^2 - (1 + \epsilon)\kappa^2\Phi^4) \right)}
{6 + \epsilon (1 + \epsilon)\kappa^2 \Phi^2}  ~.  \label{dyn3}
\eeq
Remember that \eqref{Fr1xi} constrains the initial conditions. 
Once the equations are written this way, we can apply an important theorem about the 
existence of solutions in the neighbourhood of $\epsilon=0$. 
Later on however we will return to the original form given in 
eq.\eqref{Fr1xi} - \eqref{ddPhixi}.

%
Equations \eqref{dyn1},\eqref{dyn3} are well-defined if the denominator appearing in 
these equations does not vanish, viz. 
\beq
6 +\epsilon(1 + \epsilon)\kappa^2\Phi^2 \ne 0 .\label{regular}
\eeq
Clearly, vanishing of the denominator is possible for $\epsilon<0$ only with 
$\Phi\sim \frac{\sqrt{6}}{\kappa} \frac{1}{\sqrt{|\epsilon|}}$ for $\epsilon\to 0$.
One can show (see e.g. \cite{Per} for a formal statement of the theorem) that in the region 
where the inequality \eqref{regular} holds, there exists some time interval $-a\le t\le b$ 
with $a,b>0$ where regular solutions $\Phi(t)$, $y(t)$ and $H(t)$ exist which are 
analytic functions of $\epsilon$ around $\epsilon=0$. 
This time interval can be extended as long as the solution does not diverge and 
inequality \eqref{regular} is satisfied.

As we show now, we can take $b\to \infty$ for $\epsilon$ sufficiently small.  
Indeed, using only continuity with respect to $\epsilon$ we can write for $t<b$
\beq
H(t) &=& H^{(0)}(t) + O(\epsilon)~,\label{He1}\\
\Phi(t) &=& \Phi^{(0)}(t) + O(\epsilon)~.\label{Phie1}
\eeq
We use now the crucial property that (the regular solutions) $H^{(0)},~\Phi^{(0)}$ are 
bounded functions for all times $t>0$ satisfying $\Phi^{(0)}(t) < \frac{\sqrt{6}}{\kappa},~ 
H^{(0)}(t) < \sqrt{\frac{\Lambda}{3}}$. 
Now it is clear that in the neighbourhood of the zeroth order solution, the equations are 
regular as they satisfy \eqref{regular} for all times $t>0$ so that the theorem applies.
In eq.\eqref{Phie1} the initial value should be a function of 
$\epsilon$ too with $\Phi_0\to \Phi^{(0)}_0$. We will consider in the neighbourhood 
of a given $\Phi^{(0)}$ a family of solutions depending on $\epsilon$ with a fixed initial 
value $\Phi_0$. 
In this way the initial conditions at the bounce are independent of $\epsilon$ and the function 
$\Phi^{(0)}$ to which $\Phi$ tends for $\epsilon\to 0$ is completely defined.    
A subtlety arises here for $\epsilon<0$ because some initial values leading to regular 
solutions start outside the allowed interval of initial values for $\epsilon=0$. 
We have checked numerically that solutions starting there for $\epsilon=0$ are regular 
indeed. These solutions were discarded not for mathematical but for physical reasons as they 
start with $F<0$. But from a mathematical point of view, such solutions are perfectly 
admissible.

We would like to show further that the time interval where \eqref{He1},\eqref{Phie1} apply 
can be extended to infinity. Let us assume that $\Phi(t)$ (and/or $H(t)$) diverges at some 
time $b<\infty$. Then $\Phi(t)$ (and/or $H(t)$) could become arbitrarily large for $t\to b$ 
but this would clearly be in contradiction with \eqref{He1},\eqref{Phie1}. So we have shown 
that \eqref{He1}-\eqref{Phie1} is valid also for t going to infinity. It is also obvious that 
in a neighbourhood around $\epsilon = 0$ the condition \eqref{Phimax} is satisfied so that 
the solutions obtained are physically acceptable as well. 
Note that this does not preclude the existence of regular solutions for larger 
values of $\epsilon$.

To summarize, we get a family of solutions $H(t)$ and $\Phi(t)$ arbitrarily close 
to the conformally invariant solutions $H^{(0)},~\Phi^{(0)}$. This means that conformal 
invariance, though crucial for the derivation of \emph{exact analytical} expressions,   
is not needed in order to have regular bouncing solutions. Of course, it is a priori 
not clear how far one can depart from conformal invariance. 
This will be investigated later with numerical calculations. However before resorting to 
numerical calculations, we can gain further analytical insight by studying the behaviour 
of this family of solutions at each order in a systematic expansion in $\epsilon$.   

Like for $\epsilon=0$, an analysis in the asymptotic regime \cite{BGPS15}, assuming only 
$H\to$ constant, gives that the solution with initial value $\Phi_{0,cr}$ tends asymptotically 
to $\tilde\Phi= \tilde\Phi^{(0)}~\sqrt{1+\epsilon}$, with $\Phi_{0,cr}$ depending also on 
$\epsilon$.
From \eqref{Fr1xi} we have $H(t)\to \sqrt{\frac{\Lambda}{3}}$ when $\Phi\to 0$ and also 
when $\Phi\to \tilde\Phi$. So all regular solutions tend asymptotically to a de Sitter 
space while gravity tends dynamically to GR. 

\section{Asymptotic behaviour at all order in $\epsilon$} 

We want now to investigate the solutions obtained at each order in 
a systematic expansion in $\epsilon$. In particular analytic expressions 
will be obtained at first order in the asymptotic regime $t\to \infty$. 
Using analyticity in $\epsilon$, we write 
\beq
H(t) &=& H^{(0)}(t) + \epsilon H^{(1)}(t) + O(\epsilon^2)~,\label{He2}\\
\Phi(t) &=& \Phi^{(0)}(t) + \epsilon \Phi^{(1)}(t) + O(\epsilon^2)~.\label{Phie2}
\eeq
In a standard way, substituting these expressions in our basic equations 
\eqref{Fr1xi}-\eqref{ddPhixi} with the definition \eqref{xi}, 
we obtain at first order three linear differential equations for $H^{(1)}(t)$, 
$\Phi^{(1)}(t)$ with time-dependent coefficients depending on the zeroth order 
solutions $H^{(0)},~\Phi^{(0)}$. Equation \eqref{Fr1xi} yields  
\beq
&&\left( - 6 \kappa^{-2} H^{(0)} + H^{(0)}\Phi^{(0)2} + \Phi^{(0)} \dot{\Phi}^{(0)} \right) H^{(1)} + 
\left( H^{(0)}\Phi^{(0)} + \dot{\Phi}^{(0)} \right) \dot{\Phi}^{(1)} + \nonumber \\
&&\left( H^{(0)2} \Phi^{(0)} - 4 c \Phi^{(0)3} + H^{(0)} \dot{\Phi}^{(0)} \right) \Phi^{(1)} +
3 H^{(0)2}\Phi^{(0)2} + 6H^{(0)}\Phi^{(0)} \dot{\Phi}^{(0)} = 0~. \label{B1}
\eeq
Proceeding in the same way we obtain from \eqref{Fr2xi}
\beq
&& \left( 2 \kappa^{-2} - \frac{\Phi^{(0)2}}{3} \right) \dot{H}^{(1)}  
 - \frac{1}{3}\Phi^{(0)} \ddot{\Phi}^{(1)} + \frac{1}{3}\Phi^{(0)} \dot{\Phi}^{(0)} H^{(1)} 
\nonumber \\
&+& \frac{1}{3} \left( - 2 \Phi^{(0)} \dot{H}^{(0)} + H^{(0)} \dot{\Phi}^{(0)} 
    - \ddot{\Phi}^{(0)} \right) \Phi^{(1)}  + 
\left( \frac{1}{3} H^{(0)} \Phi^{(0)} + \frac{4}{3} \dot{\Phi}^{(0)} \right) \dot{\Phi}^{(1)} 
\nonumber \\ 
&-& 2 \left( \Phi^{(0)2} \dot{H}^{(0)} - H^{(0)} \Phi^{(0)} \dot{\Phi}^{(0)} + \dot{\Phi}^{(0)2} + 
                  \Phi^{(0)} \ddot{\Phi}^{(0)} \right) = 0~,  \label{B2}
\eeq
and finally from \eqref{ddPhixi}
\begin{equation*}
\ddot{\Phi}^{(1)} + 3H^{(0)} \dot{\Phi}^{(1)} +
           \left( 2 H^{(0)2} - 12c \Phi^{(0)2} + \dot{H}^{(0)} \right) \Phi^{(1)} = 
\end{equation*}
\beq
- \left( 4H^{(0)} \Phi^{(0)} + 3 \dot{\Phi}^{(0)} \right) H^{(1)} - \Phi^{(0)} \dot{H}^{(1)} 
- 6 \left( 2 H^{(0)2} \Phi^{(0)} + \Phi^{(0)} \dot{H}^{(0)} \right)~. \label{B3}
\eeq
The right hand side of equation \eqref{B1} is a first integral of \eqref{B2}, \eqref{B3}. 
As the coefficients of the highest derivatives in \eqref{B2}, \eqref{B3} do not 
vanish, and none of the other coefficients are singular, the solutions of our linear 
system are regular. 
Using these results, we proceed with the study of the solutions at first order 
and investigate their behaviour in the asymptotic regime $t\to \infty$. 
So in what follows equations and solutions refer specifically to this regime. 
As was shown in \cite{BGPS15}, the family of lowest order solutions $\Phi^{(0)}$ 
tending to zero obey at leading order the equation
\beq
\ddot{\Phi}^{(0)} + \sqrt{3 \Lambda} \dot{\Phi}^{(0)} +\frac{2 \Lambda}{3} \Phi^{(0)} = 0~,
\label{ddPhi0as}
\eeq
whose general solution is  
\beq
\Phi^{(0)} \sim C_1 ~{\rm exp}\left( -\sqrt{\frac{\Lambda}{3}}~t \right) + 
C_2  ~{\rm exp}\left( -2 \sqrt{\frac{\Lambda}{3}}~t \right)~,
\label{Phi0as}
\eeq
where $C_1,~C_2$ are constants. Remember that there exists another unique solution 
$\Phi^{(0)}$ tending asymptotically to the nonzero finite value $\tilde{\Phi}^{(0)}$.
For $H^{(0)}(t)$ we have at leading order
\beq
H^{(0)} &\sim& \sqrt{\frac{\Lambda}{3}}~, \label{H0as}\\
\dot{H}^{(0)} &\sim&  8 \frac{\Lambda}{3} 
                {\rm exp} \left( - 4 \sqrt{\frac{\Lambda}{3}}~t \right)~.\label{dH0as}
\eeq
Using \eqref{Phi0as}, \eqref{H0as}, we see immediately from \eqref{B1} that 
we must have $H^{(1)}\to 0$ if $\Phi^{(1)}$ and $\dot{\Phi}^{(1)}$ remain bounded. 
Even if we do not assume a priori that $\Phi^{(1)}$ and $\dot{\Phi}^{(1)}$ remain bounded, 
we can use \eqref{B1} to write  
\beq
H^{(1)} = A^{(1)}(t)~\dot{\Phi}^{(1)} + B^{(1)}(t)~\Phi^{(1)} + D^{(1)}(t)~, 
~~~~~~~~~~~~~~A^{(1)},~B^{(1)},~D^{(1)}\to 0~.\label{AB}
\eeq
Therefore in this regime, \eqref{B3} becomes at leading order 
\beq
\ddot{\Phi}^{(1)} + \sqrt{3 \Lambda} ~\dot{\Phi}^{(1)} + \frac23 \Lambda \Phi^{(1)} = I^{(1)}(t)~,
\label{ddPhi1as}
\eeq
where the inhomogeneous part $I^{(1)}(t)= - 4 \Lambda \Phi^{(0)}$ decays exponentially. 
Clearly the inhomogeneous solution decays exponentially too as we will see explicitly. 
The homogeneous part of eq.\eqref{ddPhi1as} is precisely the asymptotic equation satisfied 
by $\Phi^{(0)}$. Though the inhomogeneous solution decays exponentially, it gives the solution 
of eq.\eqref{ddPhi1as} at leading order, namely  
\beq
\Phi^{(1)} \sim  - 4 C_1 \sqrt{3 \Lambda}~t 
                         ~{\rm exp}\left( -\sqrt{\frac{\Lambda}{3}}~t \right)~.
\label{Phi1as}
\eeq
As we have from \eqref{Phi1as} that $\Phi^{(1)}$ and $\dot{\Phi}^{(1)}$ vanish asymptotically, 
we know that $H^{(1)}$ goes to zero as mentioned earlier. We find again at leading order 
from \eqref{B1} using \eqref{Phi0as}
\beq
H^{(1)} \sim - \frac{C_1^2 \kappa^2}{2} \sqrt{\frac{\Lambda}{3}} 
                          {\rm exp}\left( - 2 \sqrt{\frac{\Lambda}{3}}~t \right)~.\label{H1as}
\eeq 
Hence for $\epsilon\to 0,~\epsilon > 0$, we have (asymptotically) $H\to H^{(0)}$ with 
$H < H^{(0)}$ and $\dot{H}>0$. On the other hand for $\epsilon\to 0,~\epsilon < 0$, we have 
$H\to H^{(0)}$ with $H > H^{(0)}$ and, using \eqref{dH0as} and \eqref{H1as}, $\dot{H}<0$. 
The latter behaviour is different from that found in the conformally coupled case. 
This is in agreement with our numerical simulations (see below). 

So we have derived explicit expressions for $\Phi^{(1)}$ and $H^{(1)}$ for $t\to \infty$. 
It is seen that $H^{(1)}$ tends much more rapidly to zero than $\Phi^{(1)}$. In particular 
for regular solutions, $H\to H^{(0)}$ much quickier than $\Phi\to \Phi^{(0)}$ for $\epsilon$ 
very small. Following the same reasoning at higher orders, it is easy to see that if $H^{(n-1)}$ 
and $\Phi^{(n-1)}$ tend to zero, then this will be also true for $H^{(n)}$ and $\Phi^{(n)}$. 
Indeed the constraint equation \eqref{B1} yields at $n$th order 
\beq
H^{(n)} = A^{(n)}(t)~\dot{\Phi}^{(n)} + B^{(n)}(t)~\Phi^{(n)} + D^{(n)}(t)~, 
          ~~~~~~~~~~~~~~A^{(n)},~B^{(n)},~D^{(n)}\to 0~.\label{A'B'}
\eeq
Then the homogeneous equation satisfied by $\Phi^{(n)}$ will be again similar to 
\eqref{ddPhi0as} with a nonhomogeneous part $I^{(n)}(t)$ decaying exponentially. 
To summarize, we have shown explicitly that $H^{(n)}$ and $\Phi^{(n)}$ tend exponentially 
to zero for $t\to \infty$ at all orders $n$. 

\section{Numerical simulations}

\begin{figure}
\begin{centering}
\includegraphics[scale=.7]{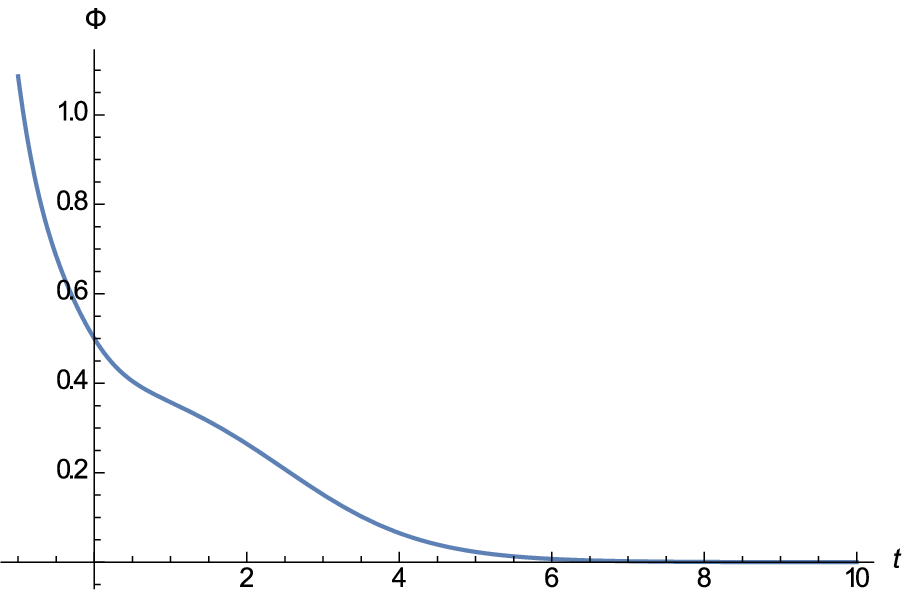}~~
\includegraphics[scale=.7]{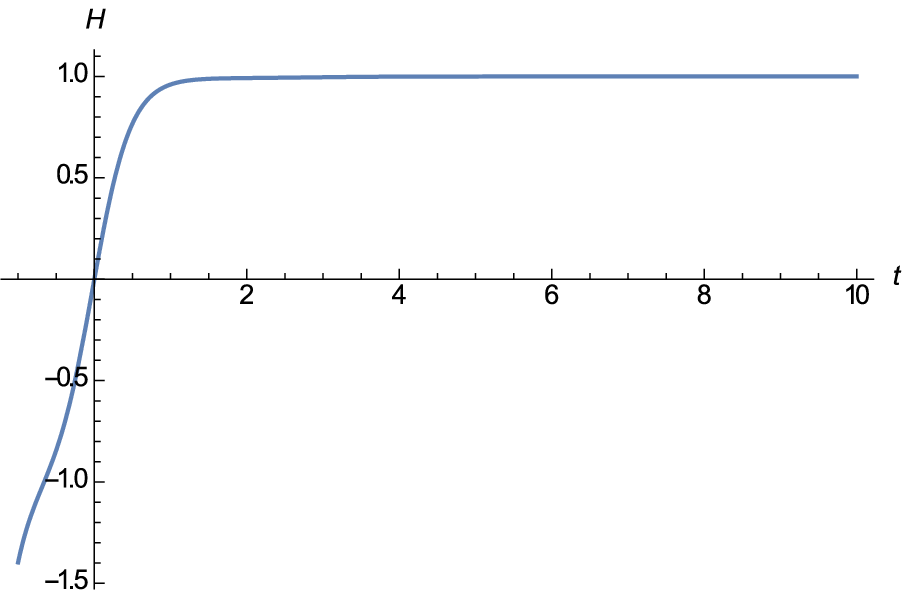}
\par\end{centering}
\caption{The functions $\Phi(t)$ and $H(t)$ are shown starting before the bounce (located at 
$t=0$) for $\epsilon=0.1$. The model parameters are $\kappa=\sqrt{20},~\Lambda=3$ and $c=3$. 
For this model we have $\Phi^{(0)}_{max}=0.5477,~\Phi_{0,min}=0.4728,~\epsilon_{max}=0.3416$ and 
$\Phi_{max} = \Phi^{(0)}_{max} \frac{1}{\sqrt{1+\epsilon}}=0.5222$. 
We find numerically $0.4807 < \Phi_{0,crit} < 0.4808$. As we have $\epsilon<\frac18$, 
$\Phi(t)$ shows no oscillations while having $\Phi_0 = 0.5$ larger than $\Phi_{0,crit}$ ensures 
its convergence. As for the conformally invariant case, $\Phi$ decreases monotonically, with 
an inflexion though, starting around $\tilde{\Phi}=0.42817$, while $H$ increases monotonically 
to $\sqrt{\frac{\Lambda}{3}}$. This behaviour of $H$ persists for $\epsilon\to 0,~\epsilon>0$, 
in agreement with \eqref{H1as}.}
\label{fig1}
\end{figure}

\begin{figure}
\begin{centering}
\includegraphics[scale=.7]{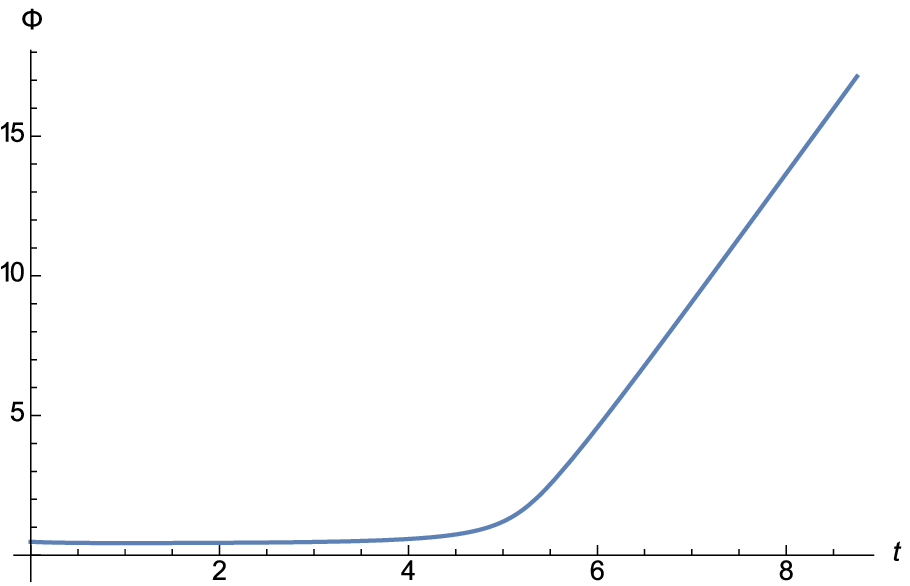}~~
\includegraphics[scale=.7]{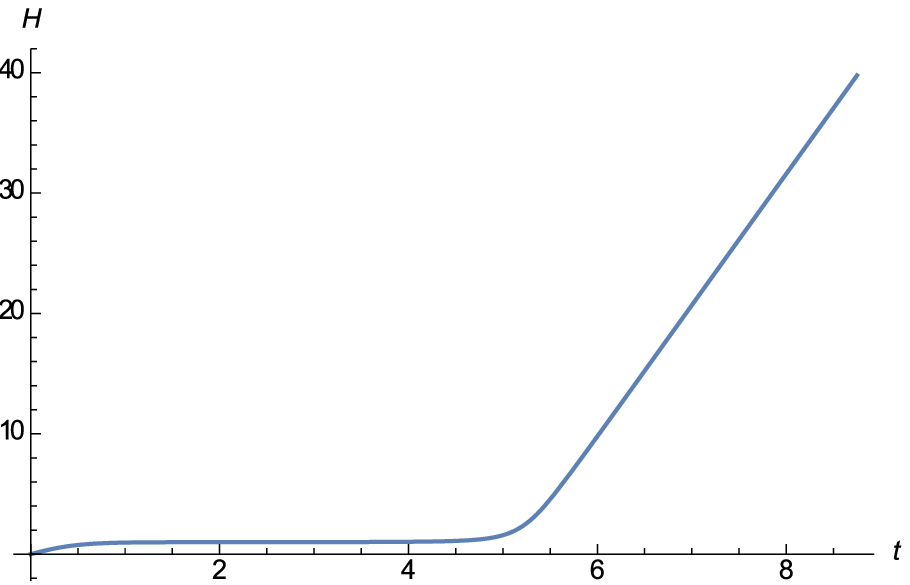}
\par\end{centering}
\caption{
The functions $\Phi(t),~H(t)$ are shown for the model parameters of Figure \eqref{fig1} and 
$\epsilon=0.1$ but with the initial condition (at the bounce) $\Phi_0= 0.48 < \Phi_{0,crit}$ 
so that $\Phi$ and $H$ diverge. In sharp contrast to the conformally invariant case, the 
divergence does not occur in a finite time. Actually, the asymptotic behaviour of both $\Phi$ 
and $H$ is linear in time according to \eqref{lin}, 
\eqref{alpha}.}
\label{fig2}
\end{figure}

\begin{figure}
\begin{centering}
\includegraphics[scale=.7]{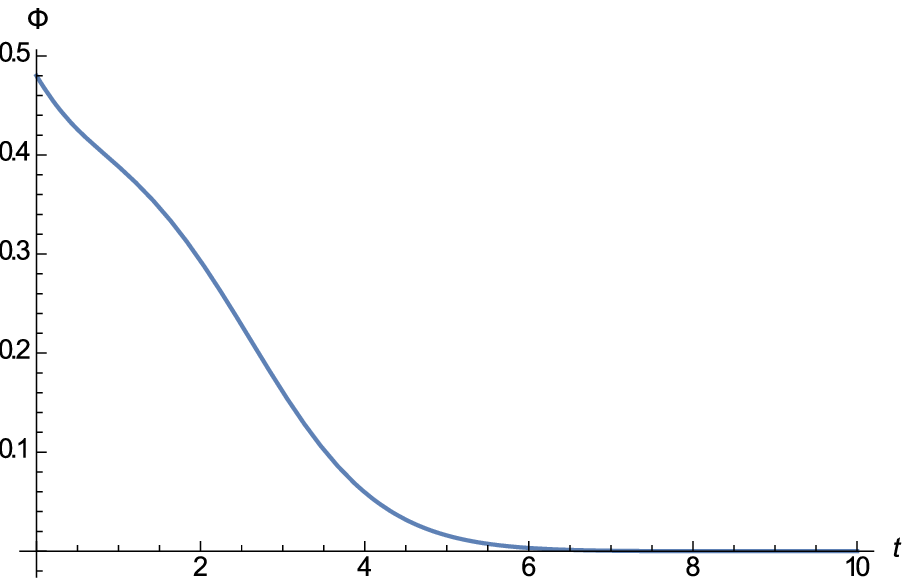}~~
\includegraphics[scale=.7]{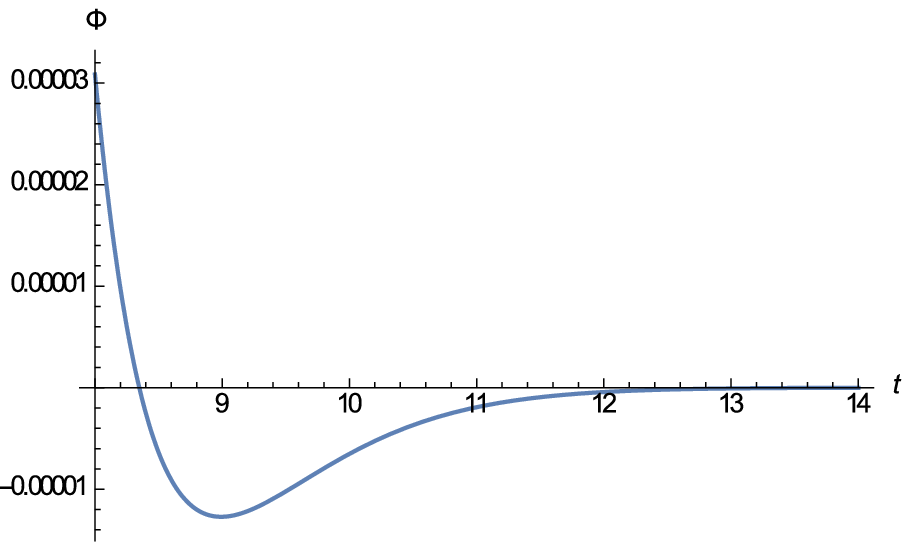}
\par\end{centering}
\caption{The function $\Phi(t)$ is shown for the model 
parameters of Figure \eqref{fig1}, but now for $\epsilon=0.25$. We find numerically 
$0.4737 < \Phi_{0,crit} < 0.4738$. We take here $\Phi_0 = 0.48$ so $\Phi$ converges. 
As we have here $\epsilon > \frac18$, oscillations must be present in $\Phi(t)$ at least 
asymptotically, but these are too small to be seen on the plot. The right panel shows a 
zoom of the left panel and we can see that $\Phi=0$ around $t=8.4$. This is 
only possible as $\dot{H}<0$ at the same time. For $\epsilon\ne 0$ one can have $\dot{H}<0$ 
but this was not possible for the conformally invariant case.} 
\label{fig3}
\end{figure}

\begin{figure}
\begin{centering}
\includegraphics[scale=.7]{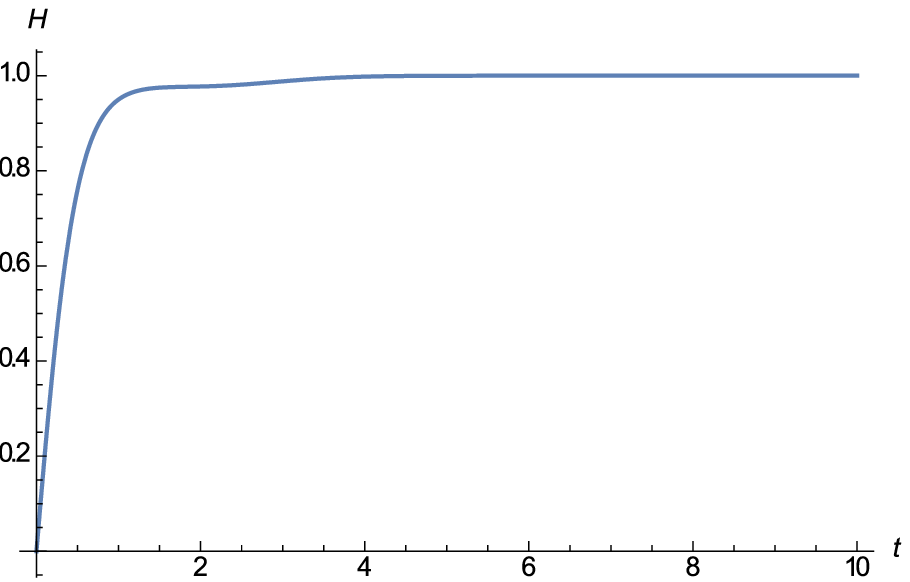}~~
\includegraphics[scale=.7]{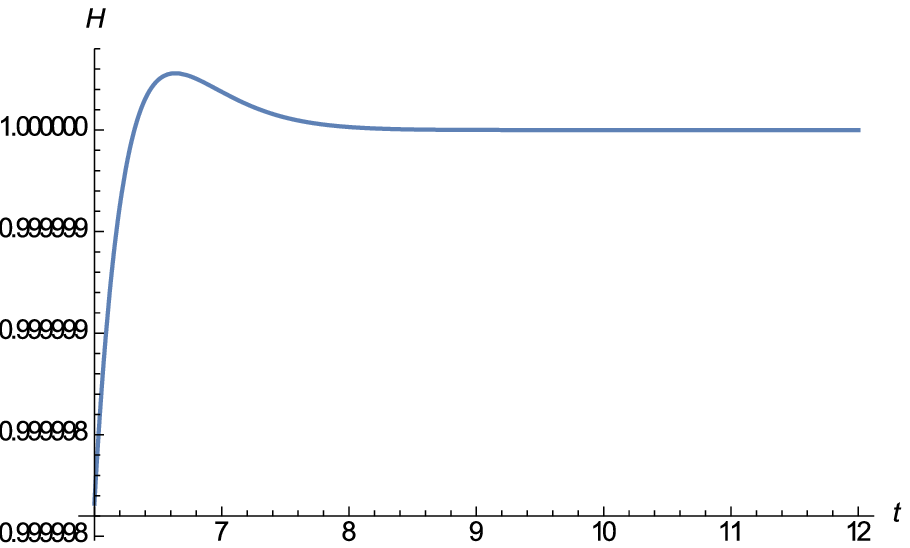}
\par\end{centering}
\caption{The function $H(t)$ is shown for the model parameters of Figure \eqref{fig1} 
and $\epsilon=0.25$.
The Hubble parameter $H$ tends to its asymptotic value but the increase is not monotonic.
A first maximum appears around $t=6.5$ as we see on the right panel. 
Actually another very weak maximum occurs around $t=12.6$. 
The asymptotic behaviour satisfies $\dot{H}<0$.
This illustrates the ability of the system to move dynamically from the 
phantom to the non-phantom regime, such a behaviour however did not occur for $\epsilon=0$ 
where a monotonic increase of $H$ was found.  
As expected, $\dot{H}<0$ at the time where $\Phi$ vanishes, see Figure \ref{fig3}.} 
\label{fig4}
\end{figure}

\subsection{$\epsilon > 0$}

We start first with $\epsilon>0$. Different cases will arise here for given system parameters 
$\Lambda,~c$ and $\kappa$. Let us take $\epsilon < \epsilon_{max}$ (see \eqref{intZ1}) so 
that a nonvanishing interval of initial values (at $t=0$) 
$\Phi_{0,min} < \Phi_0 < \Phi_{max}$ is allowed. 
As we noted already earlier, this interval of initial values is embedded in the 
allowed interval for $\epsilon=0$. 
By increasing $\epsilon$ from zero to $\epsilon_{max}$, this interval will shrink until 
it vanishes. We find that for any $\epsilon < \epsilon_{max}$ the interval of 
allowed initial values will contain a critical value $\Phi_{0,cr}$.  
Then there exists a subinterval $\Phi_{0,cr} < \Phi_0 < \Phi_{max}$ for which $\Phi$ tends 
asymptotically to zero. In that case we get for $\epsilon \le \frac18$ that $\Phi$ decreases 
monotonically to zero like $\Phi^{(0)}$.  
On the other hand for $\epsilon > \frac18$, $\Phi$ exhibits damped oscillations around zero 
in the asymptotic regime. 
Indeed, let us consider more carefully the nature of the asymptotic behaviour of $\Phi$ in 
this case. As $\Phi\to 0$ asymptotically, the equation satisfied by $\Phi$ in this 
regime is a very simple generalization of \eqref{ddPhi0as}, namely  
\beq
\ddot{\Phi} + \sqrt{3 \Lambda} ~\dot{\Phi} +\frac{2 \Lambda}{3} (1 + \epsilon) \Phi = 0~,
\label{ddPhias}
\eeq
leading to an overdamped oscillation for $\epsilon \le \frac18$. Hence 
in the neighbourhood of $\Phi^{(0)}$ the asymptotic solution shows no oscillations 
but this is no longer true for $\epsilon > \frac18$. 
An interesting point is that while the possibility to have damped oscillations is immediate 
from \eqref{ddPhias}, this is not obvious by inspection of \eqref{ddPhi1as} and of the 
similar equations for all higher order functions $\Phi^{(n)}$. 
Actually, the oscillations of $\Phi$ come from the presence of all the inhomogeneous parts 
$I^{(n)}$ as the homogeneous part alone leads to an exponentially decaying solution without 
oscillations. 
Another interesting point here is the possibility for $\Phi$ to vanish and to be negative 
in contrast to the conformally invariant case. This is because here, $\dot{H}$ can 
change sign and become negative. 
This is unlike the conformal case where $\dot{H}^{(0)}>0$, which implied by inspection 
of \eqref{Fr2xi} that $\Phi = 0$ is impossible. 
We found numerically as expected that $\Phi$ has zeroes only in regions where 
$\dot{H}<0$. Moreover we find $\dot{H}<0$ in the asymptotic regime 
for $\epsilon > \frac18$ so that an infinite number of zeroes of $\Phi$ is possible. 
Actually $\dot{H}<0$ can already occur for $\epsilon < \frac18$.  
If we start with $\Phi_0 < \Phi_{0,cr}$, the solution will diverge for $t\to \infty$. 
However it is interesting that they do not become singular in a finite time as 
for $\epsilon=0$. Instead both $\Phi$ and $H$ tend to infinity for $t\to \infty$. 
It is not hard to see from equations \eqref{Fr1xi} - \eqref{ddPhixi} that 
\beq
\Phi\sim \alpha t~,~~~~~~~~~~~~~~~~~H\sim \beta t~\label{lin} 
\eeq
are possible asymptotic solutions for $t\to \infty$. Their ratio is constant in 
this regime and we find easily  
\beq
\frac{\alpha}{\beta} = \sqrt{\frac{1 + \epsilon}{2c}}~. \label{alpha}
\eeq
We find numerically that both slopes go to infinity as $\epsilon$ goes to zero.
Remember that in the conformal case the system has the remarkable property that the dynamics 
of $\Phi$ decouples essentially from $H$, more precisely in equation \eqref{Fr1} 
only a first integral of eq.\eqref{ddPhi} will appear. 
Hence in that case one could have that $\Phi$ diverges, in a finite time even, while $H$ 
remains perfectly regular. This is no longer true here. 
Figures \ref{fig1} - \ref{fig4} show these different behaviours for a specific choice 
of the system parameters $\kappa$, $\Lambda$ and $c$ and $\epsilon=0.1,~0.25$. The 
qualitative behaviour is similar for other system parameters.  

\subsection{$\epsilon < 0$}

Let us turn now our attention to $\epsilon<0$. 
In contrast to the case with $\epsilon>0$, now the interval of allowed initial values 
includes the interval for $\epsilon=0$, so some initial values are possible which were 
forbidden in the conformally invariant case. 
It is further clear from eq.\eqref{ddPhias} which is valid for any sign of $\epsilon$ 
that $\Phi$ behaves asymptotically like an overdamped oscillation. 
A value $\Phi_{0,crit}$ is found again and interestingly we find that $\Phi_{0,crit}$ itself 
can be larger than $\Phi^{(0)}_{max}$. 
For $\Phi_0 > \Phi_{0,crit}$, $\Phi$ decreases monotonically and tends 
asymptotically to zero. The Hubble parameter $H$ in turn increases first, before 
decreasing asymptotically and tending to $\sqrt{\frac{\Lambda}{3}}$. 
For $\epsilon\to 0$, $H$ will first reach a maximum before tending to its 
asymptotic value with $\dot{H}<0$. This is in perfect agreement with our expression 
\eqref{H1as}: asymptotically, we have $H > H^{(0)}$ and $\dot{H}<0$ (while of course 
$H\to H^{(0)}$ and $\dot{H}\to 0$).

When $\Phi_0 < \Phi_{0,crit}$, $\Phi$ is found to diverge in a finite time. 
This is similar to the singular behaviour occuring for $\epsilon=0$. 
Now however $H\to -\infty$ in a finite time as well, corresponding to a Big Crunch.     
 
For $\epsilon<0$ a critical value $\epsilon_c$ is also found such that for 
$|\epsilon| > |\epsilon_c|$ both $\Phi$ and $H$ diverge for whatever value $\Phi_0$ 
in the allowed interval of initial values. However for $\epsilon<0$ this divergence 
happens in a finite time, in a way similar for $|\epsilon| < |\epsilon_c|$ with 
$\Phi_0 < \Phi_{0,crit}$.

\begin{figure}
\begin{centering}
\includegraphics[scale=.7]{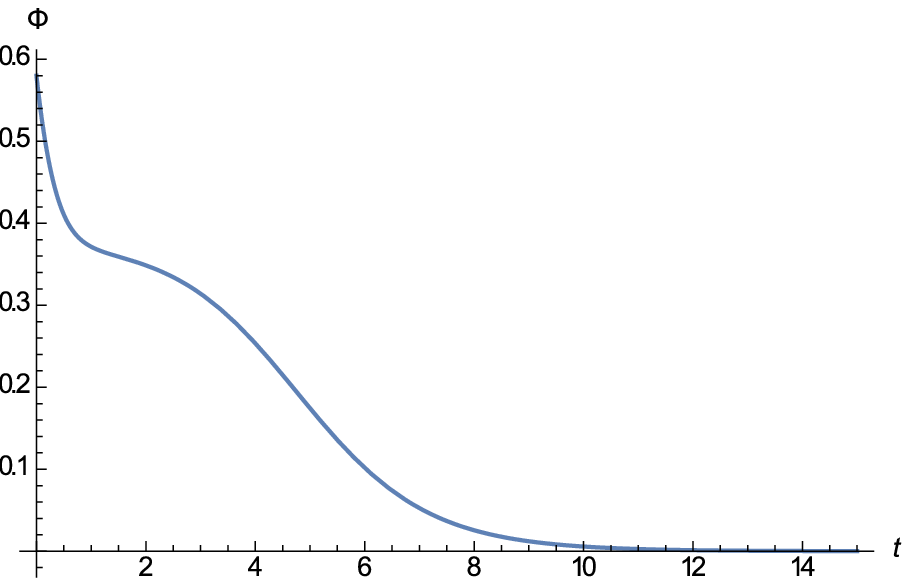}~~
\includegraphics[scale=.7]{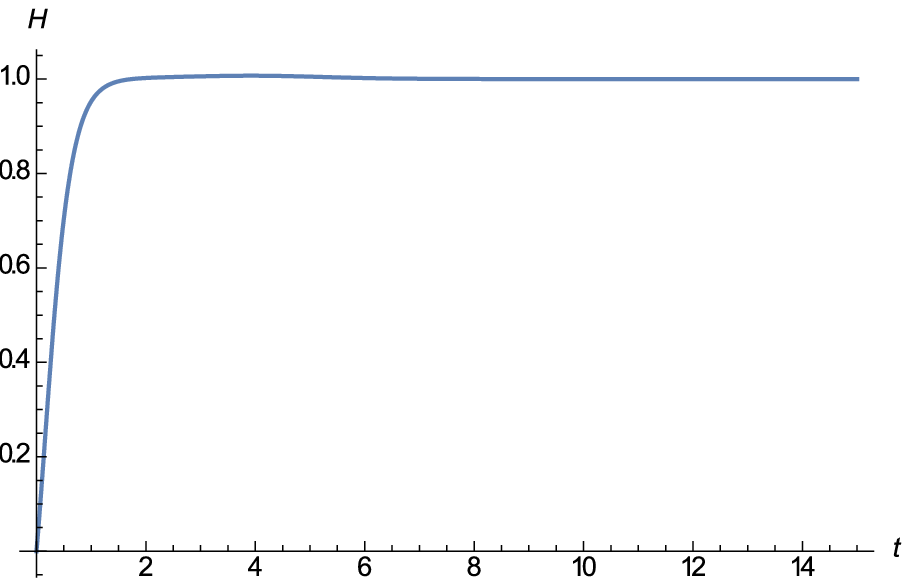}
\par\end{centering}
\caption{The functions $\Phi(t)$ and $H(t)$ are shown starting for the model 
parameters of Figure \ref{fig1} and $\epsilon_c < \epsilon = -0.15$ with 
$-0.177 < \epsilon_c < -0.176$. For $|\epsilon| > |\epsilon_c|$ both functions would 
diverge for any intial value $\Phi_0$. We find $0.559 < \Phi_{0,crit} < 0.560$ and we 
take $\Phi_0 = 0.58 > \Phi_{0,crit}$. 
The behaviour is essentially similar to the conformally invariant case, the increase of 
$H$ however is not monotonic as will be seen on Figure \ref{fig6}. }
\label{fig5}
\end{figure}

\begin{figure}
\begin{centering}
\includegraphics[scale=.7]{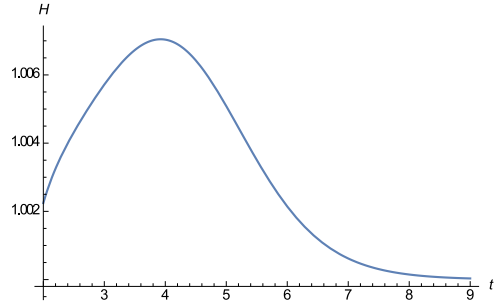}~~
\par\end{centering}
\caption{The function $H(t)$ is shown, it is a zoom of Figure 
\ref{fig5} showing its nonmonotonic increase. 
The system evolves dynamically from a phantom to a non phantom regime. 
Actually this behaviour persists for $\epsilon\to 0,~\epsilon < 0$, 
though the maximum becomes less and less pronounced, hence $H\to H^{(0)}$ from 
above with $\dot{H}<0$, in perfect agreement with eq.\eqref{H1as}.}
\label{fig6}
\end{figure}

\begin{figure}
\begin{centering}
\includegraphics[scale=.7]{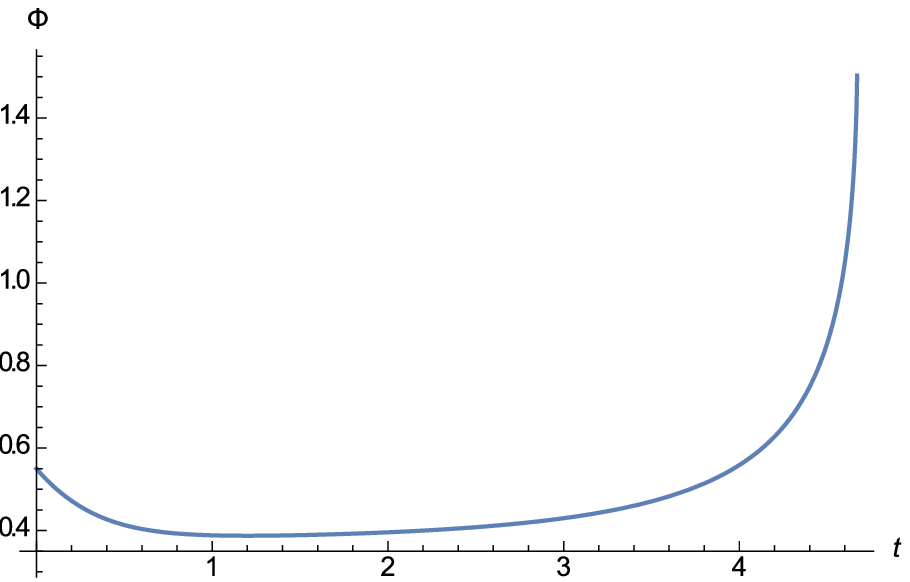}~~
\includegraphics[scale=.7]{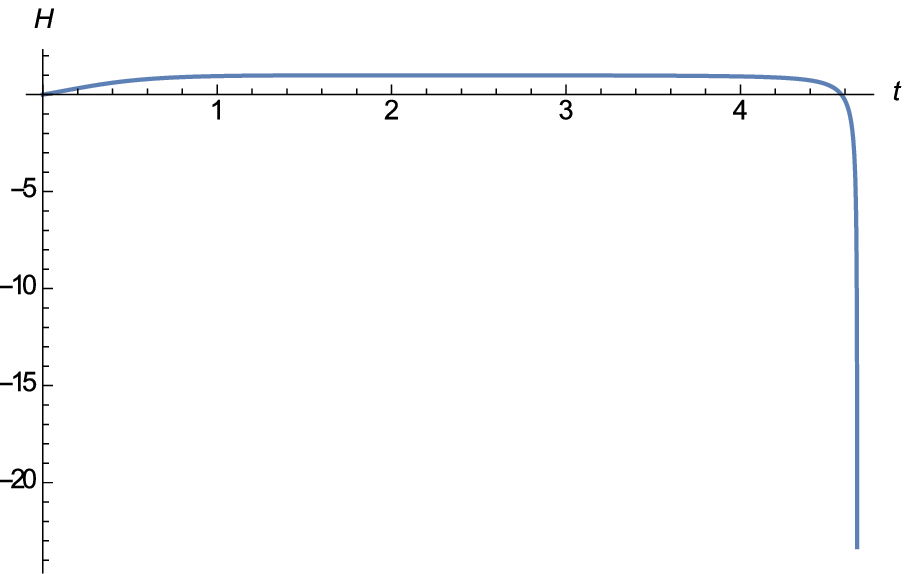}
\par\end{centering}
\caption{The functions $\Phi(t)$ and $H(t)$ for the model parameters of Figure \ref{fig1} 
for $\epsilon=-0.15$ and $\Phi_0 = 0.55 < \Phi_{0,crit}$. Now both $\Phi$ and $H$ diverge in 
a finite time. The divergence of $\Phi$ is similar of that found for $\epsilon=0$. 
The divergence of $H$ which is impossible for $\epsilon=0$, represents a contracting 
universe with a Big Crunch in the unphysical regime $F<0$.}
\label{fig7}
\end{figure}

Figures \ref{fig5} - \ref{fig7} show these various behaviours for specific values 
of the system parameters. The same qualitative behaviour is obtained for other system 
parameters.

\section{Conclusions}

A family of spatially-flat bouncing solutions with a nonzero measure set of initial 
conditions regular after the bounce (located at $t=0$) were recently found in the framework 
of scalar-tensor gravity. In this model, the scalar field was conformally coupled, allowing 
the derivation of exact analytical expressions. A crucial aspect was the decoupling of the 
Hubble parameter $H(t)$ from $\Phi(t)$. Indeed $H(t)$ coupled only to a first integral of 
$\Phi(t)$ and does not depend on the dynamics of $\Phi(t)$. 
The primary purpose of the present work is to extend these results and to investigate 
whether regular bouncing solutions exist also for 
$\xi\equiv \frac16 (1 + \epsilon)\ne \frac16$. 

We have shown that there exists a family of regular bouncing solutions 
at least in the neighbourhood of the conformally coupled ones, corresponding to 
$\xi\ne \frac16$ arbitrarily close to $\frac16$.  
These scalar field solutions vanish exponentially with damped oscillations and an infinite 
number of zeroes when $\frac18 < \epsilon < \epsilon_{max}$, and in a monotonous way
for $0 < \epsilon \le \frac18$. Regular non oscillating solutions tending to zero were also 
found when $\epsilon < 0$. 
In a systematic expansion of the solutions in terms of $\epsilon$ we show explicitly that 
the solutions $\Phi$ and $H$ vanish exponentially at each order. Analytic asymptotic 
expressions were also given at first order.  

Proceeding with a numerical study, all the above properties were confirmed also for significant 
departures from $\epsilon=0$. 
Like in the conformally coupled case, whenever regular solutions exist we find 
that part of the allowed initial interval with $\Phi_0 > \Phi_{0,cr}$ will lead to regular 
solutions, while the other initial values will lead to divergent solutions. 

For $\epsilon > 0$ we find that regular bouncing solutions exist for any 
$0 < \epsilon < \epsilon_{max}$. 
As $\epsilon\to \epsilon_{max}$ and hence $\Phi_{max}\to \Phi_{0,min}$, 
we always find a critical initial value $\Phi_{0,min}<\Phi_{0,crit}<\Phi_{max}$. 
The divergence of both $\Phi$ and $H$ for $\Phi_0 < \Phi_{0,crit}$ is linear in time 
for $t\to \infty$ in contrast with the conformally coupled case where $\Phi$ 
diverges in a finite time. Here too it implies in particular that the model becomes 
unviable in a finite time when $F$ becomes negative.  

When $\epsilon<0$, a critical value $\epsilon_c$ was found such that all 
solutions diverge for $\epsilon < \epsilon_c$. For $\epsilon_c < \epsilon < 0$ however, a 
critical initial value $\Phi_{0,min}<\Phi_{0,crit}<\Phi_{max}$ and regular solutions for 
$\Phi_0 > \Phi_{0,crit}$ are found. In this case, the nonregular solutions diverge in a 
finite time with the occurrence of a Big Crunch.   

Another aspect of our work is related to the metric dynamics. 
As for the conformally invariant case, all regular bouncing solutions tend asymptotically 
to a de Sitter space with Hubble parameter $\sqrt{\frac{\Lambda}{3}}$ while gravity tends to 
GR. However in contrast to the conformally coupled case where $H(t)$ does not depend on 
$\Phi^{(0)}(t)$ (for given parameters of the system) -- a remarkable property allowing a 
complete integration of the problem -- this is no longer the case when $\epsilon\ne 0$.
To incorporate this bouncing model in a realistic cosmology, some dynamical evolution into a 
decelerated stage after the bounce should take place. Though we find various cases where 
$\dot{H}$ becomes negative, this change is not significant enough and more drastic 
departures from conformal invariance are presumably needed.   



\end{document}